\documentclass[12pt]{article}
\pdfoutput=1
\usepackage{amssymb,graphicx}
\usepackage{epstopdf}
\usepackage{amsmath,amsfonts}
\usepackage{epsfig,color}

\begin{document}

\sloppy

\sloppy

\title{\bf Caustic free completion of\\ pressureless perfect fluid and k-essence}
\author{Eugeny Babichev$^{a,b}$\footnote{{\bf e-mails}:
   eugeny.babichev@th.u-psud.fr, sabir.ramazanov@gssi.infn.it}\ ~and Sabir Ramazanov$^{c}$\\
 \small{$^a$\em Laboratoire de Physique Th\'eorique, CNRS,} \\ 
 \small{\em  Univ. Paris-Sud, Universit\'e Paris-Saclay, 91405 Orsay, France}\\
 \small{$^b$ UPMC-CNRS, UMR7095, Institut d'Astrophysique de Paris,}\\
\small{ ${\mathcal{G}}{\mathbb{R}}\varepsilon{\mathbb{C}}{\mathcal{O}}$,
98bis boulevard Arago, F-75014 Paris, France}\\
  \small{$^c$\em Gran Sasso Science Institute (INFN), Viale Francesco Crispi 7, I-67100 L'Aquila, Italy}\\
 }

{\let\newpage\relax\maketitle}

\begin{abstract}
Both k-essence and the pressureless perfect fluid develop 
caustic singularities at finite time. We further explore the connection 
between the two and show that they belong to the same 
class of models, which admits the caustic free completion 
by means of the canonical complex scalar field. Specifically, the free massive/self-interacting complex scalar reproduces dynamics 
of pressureless perfect fluid/shift-symmetric k-essence under certain initial conditions in the limit of large mass/sharp self-interacting potential. 
We elucidate a mechanism 
of resolving caustic singularities in the complete picture. The collapse time is promoted to complex number. 
Hence, the singularity is not developed in real time. The same conclusion holds for 
a collection of collisionless particles modelled by means of the Schroedinger equation, or ultra-light axions (generically, coherent oscillations of bosons in the Bose--Einstein condensate state). 

\end{abstract}

\section{Introduction}
Theories with non-canonical kinetic terms are of particular interest 
in cosmology and in the field of modified gravities. These include k-essence~\cite{ArmendarizPicon:1999rj, ArmendarizPicon:2000dh, ArmendarizPicon:2000ah, Garriga:1999vw}, 
ghost condensate~\cite{ArkaniHamed:2003uy} and Generalized Galileons, or Horndeski, models~\cite{Horndeski:1974wa, Fairlie:1991qe, Nicolis:2008in, Deffayet:2009wt, Deffayet:2009mn, Charmousis:2011bf}. See the review~\cite{Clifton:2011jh} for a complete list. Admitting 
such non-trivial field theories opens up new prospectives for addressing standard cosmological problems, e.g., the smallness of $\Lambda$-constant and 
the initial singularity problem. 

At the same time, theories of that kind, even free of any obvious pathologies, 
may secretly possess some unappealing properties. First, they typically exhibit the sub-/superluminality and, therefore, look quite uncommon for 
a particle physicist. This is, e.g., the case of k-essence/Generalized Galileons. 
While superluminality does not immediately entail causal paradoxes~\cite{Babichev:2007dw}, it may nevertheless obstruct the UV-completion by means of a local Lorentz-invariant quantum field theory~\cite{Adams:2006sv}.  

Another shortcoming of k-essence has been revealed recently~\cite{Babichev:2016hys}: it leads to the appearance 
of caustic singularities (see also Refs.~\cite{deRham:2016ged, Mukohyama:2016ipl, Tanahashi}). 
Namely, characteristics of equations of motion cross at some finite time, what results into multivalued derivatives of the 
k-essence field. This has been proved in Ref.~\cite{Babichev:2016hys} for the case of the generic simple wave. The same conclusion holds in the class of Generalized Galileon models involving k-essence.
 
Appearance of caustics brings together k-essence and the pressureless perfect fluid (PPF). Typically employed as an approximate description for the collection of collisionless 
particles, PPF also develops singularities. There is no actual problem from the particle physics point of view: shell-crossing merely 
signals the breakdown of the fluid description. Namely, a singularity is avoided by allowing the multi-stream regime.  However, PPF is more generic and may arise as the field-theoretical construction in some 
modifications of gravity, e.g., scalar Born--Infeld theories~\cite{Felder:2002sv}, mimetic matter scenario~\cite{Chamseddine:2013kea, Chamseddine:2014vna} and the projectable version of the 
Horava--Lifshitz gravity~\cite{Horava:2009uw, Blas:2009yd}. In that case, one should design another way of curing singularities.

In the present paper, we point out an even deeper connection between the shift-symmetric k-essence and PPF. We show that they belong to the same 
class of models, which admits the caustic free completion 
by means of the canonical complex scalar field, see Section~2.
In particular, PPF corresponds to the free massive field. 
On the other hand, the generic potential with self-interactions leads to k-essence.

In what follows, we restrict to the classical level analysis. Therefore the caustic free completion discussed here should 
not be confused with the quantum field theoretical UV-completion. Even within a purely classical approach, the correspondence between the k-essence/PPF and the complex scalar is not straightforward. This clearly follows 
from the degree of freedom (DOF) counting: only one DOF is enough to describe dynamics of k-essence/PPF, 
and two DOFs are required in the case of the complex scalar. 
The problem can be addressed by a proper choice of the initial 
configuration for the complex field, as is shown in Fig.~\ref{potentials}. Modulo the cosmological drag, we set its amplitude to the constant value in the early Universe. 
This requirement fixes the frequency dependence of the field and effectively 
eliminates one extra degree of freedom. 

Using a particular example of PPF, we show how the free massive complex field reproduces its dynamics. 
Both follow the same evolution down to the times, when caustic singularities 
start to be formed. Since this point on, the discrepancy between two scenarios 
becomes unavoidable: while the description in terms of PPF breaks down, the 
actual singularity does not emerge in the complete picture. In the latter case, the collapse time is promoted to the complex quantity. Hence, the real time evolution always remains smooth. 
While we deduce this conclusion from the study of PPF evolution, 
we conjecture that the same mechanism is generic and also works 
for k-essence.



Note that the similarity between the complex scalar dynamics and superfluids is 
well-known~\cite{Son:2000ht, Son:2002zn}. In the context of k-essence models, the correspondence was 
pointed out in Refs.~\cite{Bilic:2008zk, Bilic:2008zz}. However, the idea has been barely 
used in the field of modified gravity to address shortcomings of k-essence. 
We fill in this gap in the present work. 

Furthermore, PPF can be modeled by means of the complex field---quantum mechanical wave function 
obeying Schroedinger equation~\cite{Widrow:1993qq, Widrow:1996eq, Davies:1996kp, Coles:2002sj, Short:2006md, Short:2006me, Uhlemann:2014npa}. This observation is often used as an efficient tool to study the 
dynamics of collisionless dust particles without resorting to cumbersome 
N-body simulations. Notably, in some situations of interest in cosmology, the Schroedinger equation provides the genuine description 
of the physical system. This is, e.g., the case of ultra-light axions at distances smaller than their de Broglie wavelength, or, more generally, 
bosons in the Bose-Einstein condensate state~\cite{Press:1989id, Sin:1992bg, Ji:1994xh, Lee:1995af, Hu:2000ke, Goodman:2000tg, Nucamendi:2000jw, Alcubierre:2001ea, Arbey:2003sj, Boehmer:2007um, Woo:2008nn, Chavanis:2011uv, Suarez1, Li, Hui:2016ltb}\footnote{Strictly 
speaking, the Schroedinger equation is suitable for describing the non-interacting axion. Once self-interactions are included into the analysis, it must be replaced by the Gross--Pitaevskii equation.}. 
Results of the present work can be readily applied to all those cases.

 The outline of the paper is as follows. In Section~2, building on the proposal 
of Ref.~\cite{Babichev:2016hys}, we consider the class of models, which comprises k-essence and PPF. 
In Section~3, we elaborate the conditions, under which the free massive scalar complex field reproduces the dynamics 
of PPF. There we also discuss the mechanism of the caustic avoidance.  In Section~4, we generalize the conclusions made in the context of PPF to k-essence. 
We finalize with some discussions in Section~5.

\section{Generalities}

We start with the action given by~\cite{Babichev:2016hys},  
\begin{equation}
\label{actionbasic}
S=\int d^4 x \sqrt{-g} \left[\frac{\epsilon}{2} (\partial_{\mu} \lambda)^2 +\frac{\lambda^2}{2} (\partial_{\mu} \varphi)^2 -V(\lambda) \right] \; .
\end{equation}
Here $\lambda$ and $\varphi$ are scalar fields, and $\epsilon$ is the dimensionful constant. Note that Ref.~\cite{Babichev:2016hys} deals 
with the kinetic term $\sim \lambda (\partial_{\mu} \varphi)^2$ for the field $\varphi$. With our choice~\eqref{actionbasic}, the kinetic 
term is manifestly positively defined, and we avoid any possible issues with ghost instabilities. Equations of motion following from the action~\eqref{actionbasic} are given by 
\begin{equation}
\label{lambdaeq}
\epsilon \square \lambda -\lambda (\partial_{\mu}  \varphi)^2+V' (\lambda)=0 \; .
\end{equation}
and 
\begin{equation}
\label{varphieq}
\partial_{\mu} (\sqrt{-g}\lambda^2 \partial^{\mu}  \varphi)=0 \; .
\end{equation}
Upon setting $\epsilon =0$, Eq.~\eqref{lambdaeq} reduces to an algebraic equation, which can be used
 to express the variable $\lambda$ as the function of $X\equiv g^{\mu \nu} \partial_{\mu} \varphi \partial_{\nu} \varphi$, i.e., $\lambda=F(X)$. 
Substituting this back into the action, we reproduce the shift-symmetric k-essence action, 
\begin{equation} 
\nonumber 
S=\int d^4 x \sqrt{-g} {\cal L} (X) \; ,
\end{equation}
where one should identify $\sqrt{{\cal L}' (X)}=F(X)$. 

Note that the model~\eqref{actionbasic} contains not just the k-essence. Indeed, consider the quadratic potential 
$V(\lambda)=\frac{\lambda^2}{2}$. In that case, the field $\lambda$ plays the role of the Lagrange multiplier and, hence, 
cannot be expressed as the function of $X$. Therefore, this case does not match any of k-essence scenarios. 
Still, it represents a physically relevant situation. Indeed, the stress-energy tensor for the choice $V(\lambda)=\frac{\lambda^2}{2}$ is given by,
\begin{equation}
\nonumber
T_{\mu \nu}=\lambda^2 \partial_{\mu} \varphi \partial_{\nu} \varphi \; .
\end{equation}
We recognize the pressureless perfect fluid (PPF) described by the energy density $\lambda^2$ and 
the velocity potential $\varphi$~\cite{Lim:2010yk}. PPF is perhaps the best known example of the system developing caustic singularities. We conclude that the 
k-essence and PPF indeed represent particular examples of one and the same model. 

Switching to the case of the non-zero parameter $\epsilon$, i.e., $\epsilon \neq 0$, promotes the field $\lambda$ to the dynamical degree of freedom. 
Let us elucidate the physical content of the model in that case.  For this purpose, it is convenient to redefine the variables,
\begin{equation}
\nonumber
\sqrt{\epsilon} \lambda =\tilde{\lambda}   \qquad  \tilde{\varphi}=\frac{\varphi}{\sqrt{\epsilon}} \; ,
\end{equation}
and to introduce the complex scalar field
\begin{equation}
\label{complexdef}
\Psi \equiv \Psi_1+i\Psi_2=\tilde{\lambda} \cdot e^{i \tilde{\varphi}} \; .
\end{equation}
It is easy to see that the action~\eqref{actionbasic} can be recast in the following simple form, 
\begin{equation}
\label{complexscalar}
S=\int d^4 x \sqrt{-g} \left[\frac{1}{2} |\partial_{\mu} \Psi |^2 -V(\Psi) \right]\; .
\end{equation}
Remarkably, we arrive at the action of the canonical complex scalar field.

Consider the renormalizable potential $V$ of the form, 
\begin{equation}
\label{potentialgen}
V(\Psi)=\frac{\alpha \cdot  M^2 |\Psi|^2}{2}+\frac{\beta M^4|\Psi|^4}{4\Lambda^4} \; ,
\end{equation}
where we introduced the notation
\begin{equation}
\nonumber 
M^2 \equiv \frac{1}{\epsilon} \; ;
\end{equation}
$\alpha$, $\beta$ and $\Lambda$ are some arbitrary constants. We see that the limit 
of infinitely small $\epsilon$ corresponds to infinitely heavy mass of the field $\Psi$ and/or a very steep potential (for the fixed parameters $\alpha$, $\beta$ and $\Lambda$). 

Note in passing that the complex scalar with the potential~\eqref{potentialgen} often arises in the context of Bose-Einstein condensates. That is, for a sufficiently large de Broglie wavelength and/or at large densities, the classical approach breaks down. In that regime, excitations of bosons 
are described by the complex field satisfying the Schroedinger equation (if bosons are non-interacting) or 
the Gross--Pitaevskii equation (provided that there is the self-interaction). The action~\eqref{complexscalar} can be viewed as 
the relativistic completion of the Bose--Einstein condensate. In what follows, we choose to do not concretize the physical origin of the field $\Psi$.

Let us list the models 
associated with different values of the parameters $\alpha$ and $\beta$.
\begin{itemize}
\item Setting $\beta=0$ we get a model of a free massive complex field. As it has been explained previously, this case corresponds to PPF. 
\item The spontaneous symmetry breaking potential, i.e., with $\alpha<0$ and $\beta>0$, stands for the 
simplest subluminal k-essence scenario $\sim X+X^2$,~\cite{Bilic:2008zk, Bilic:2008zz}.  
\item The unbounded potential, i.e., $\alpha>0$ and $\beta<0$, corresponds to the superluminal
k-essence model $\sim X -X^2$,~\cite{Bilic:2008zk, Bilic:2008zz}. The instability can be avoided, if we add higher powers of the field $\Psi$.
\item Finally, for $\alpha>0$ and $\beta>0$ we are led to the action $\sim (X-1)^2$~\cite{Bilic:2008zk, Bilic:2008zz},---ghost condensate. 
Not surprisingly, caustic singularities have been identified in this context as well~\cite{ArkaniHamed:2005gu}.  
\end{itemize}

Needless to say, the canonical complex scalar is free of caustic singularities. 
Nevertheless, it is unclear, how the dynamics of PPF or k-essence is reproduced in this context. The question is non-trivial, 
since dynamics of the complex scalar 
is described by two degrees of freedom, while the evolution of k-essence and PPF is fixed 
by one. Hence, the correspondence holds only for the specific initial configuration 
of the complex field. 

To paraphrase, the model of the complex scalar does not provide with the genuine completion of k-essence/PPF, 
but rather approximates the latter under certain conditions. The accuracy of the approximation 
grows with the mass parameter $M$. So, in the limit $M\rightarrow \infty$ the complex scalar exactly reproduces 
the dynamics of k-essence/PPF down to the times, when caustics are supposed to be formed. This we show in the next Section using 
the example of PPF. It is also important to stress that our discussion concerns only classical dynamics. This further precludes the direct comparison of our results with those of Ref.~\cite{Adams:2006sv}, which
 strongly indicate problems with the UV-completion of a superluminal model (e.g., versions of k-essence) in the 
framework of a local Lorentz-invariant field theory. In other words, the canonical complex scalar and (superluminal) k-essence do not match each other at the quantum level. Still, they do so at the classical level. 
Hence, replacing k-essence by the complex scalar, one keeps all the essentials of large-scale dynamics and simultaneously avoids possible issues with the putative UV-completion. 

\section{Pressureless perfect fluid}

In the bulk of the present paper, we will deal with PPF. The reason is that it corresponds to the tractable case of the free massive scalar. 
Still, the main statements formulated in this Section appear to be generic 
and can be extrapolated to the situation with the self-interacting scalar and, hence, k-essence. 

We will focus on two limiting cases: i) the homogeneous evolution of the complex scalar is dominated 
by the cosmological expansion; ii) inhomogeneities of the field $\Psi$ are large, so that the cosmic drag can be neglected. 
To simplify our analysis, we will switch off metric perturbations. This is 
justified, since caustic singularities represent the intrinsic property of PPF, i.e., they occur even in the 
absence of gravity. Other simplifications will be discussed, where relevant.

\subsection{Homogeneous case}

 \begin{figure}[tb!]
\begin{center}
\includegraphics[width=0.49\columnwidth,angle=0]{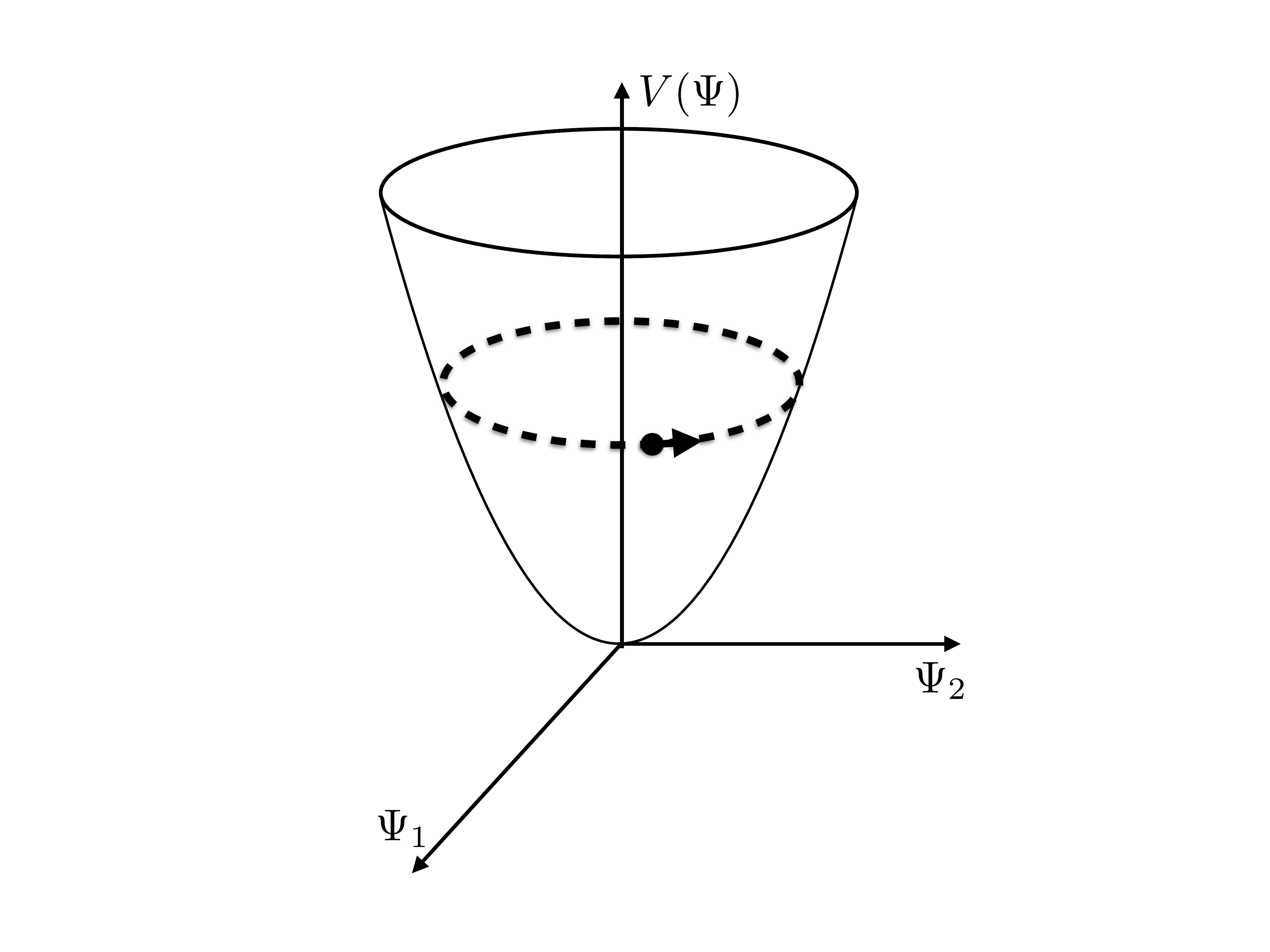}
\includegraphics[width=0.49\columnwidth,angle=0]{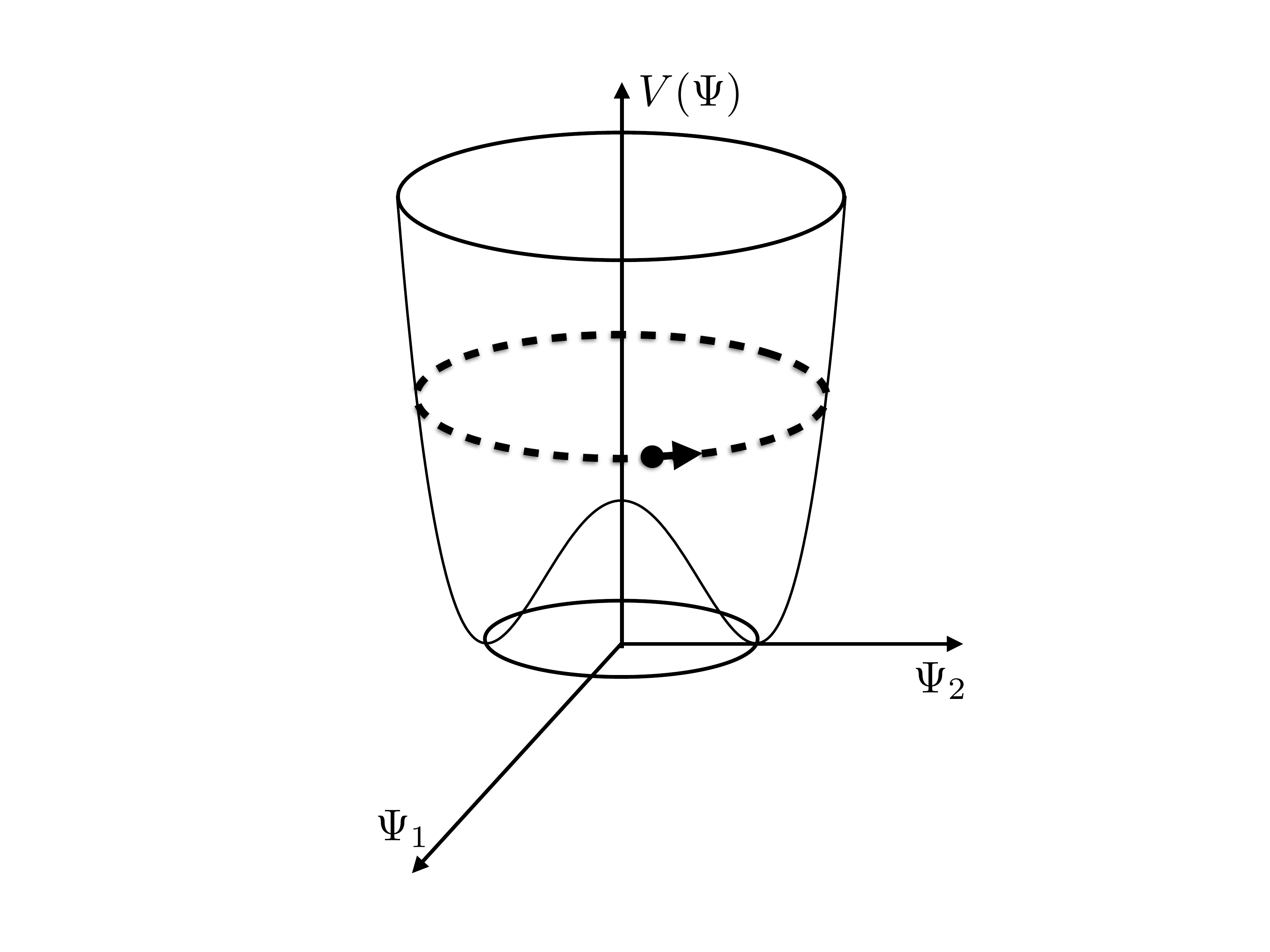}
\caption{Complex field $\Psi$ potential~\eqref{potentialgen} for the choice of parameters ($\alpha>0, \beta=0$) (free massive case, left) and ($\alpha<0, \beta>0$) (potential with spontaneous symmetry breaking, right). 
The former corresponds to PPF, while the latter stands for the simplest model of k-esseence with sub-luminality, ${\cal L} \propto X+X^2$. Dynamics of PPF 
and k-essence is reproduced for the particular configuration of the complex field depicted by the dashed line.}\label{potentials}
\end{center}
\end{figure}

In the homogeneous case, the equation of motion for the free massive field $\Psi$ (we set $\alpha=1$ and $\beta=0$ in the potential~\eqref{potentialgen}) is given by, 
\begin{equation}
\label{eqdust}
\ddot{\Psi} +3H\dot{\Psi}+M^2\Psi=0 \; .
\end{equation}
The solution to this equation reads
\begin{equation}
\label{osc}
\Psi=\frac{A}{a^{3/2}} e^{iMt}+\frac{B}{a^{3/2}} e^{-iMt} \; ;
\end{equation}
here $A$ and $B$ are some constant amplitudes. More precisely, the solution~\eqref{osc} satisfies Eq.~\eqref{eqdust} modulo terms suppressed by the ratio $\frac{H^2}{M^2}$, where 
$H$ is the Hubble rate. For the generic amplitudes $A$ and $B$, the evolution of the field 
$\Psi$ is represented by a peculiar curve\footnote{This curve is an ellipse in the limit of the static Universe.} in the configuration space. The curve reduces to the 
circle in the particular case,  
\begin{equation}
\label{initial}
B=0 \; .
\end{equation}
(alternatively, one could set $A=0$), see Fig.~\ref{potentials}. The relevance of that condition is clear. Once it is imposed, the amplitude of the field $\Psi$ is a slowly varying function (constant modulo cosmic drag). Hence, one can neglect the 
first term on the l.h.s. of Eq.~\eqref{lambdaeq}, which reduces to the constraint equation $\partial_{\mu} \varphi \partial^{\mu} \varphi=1$. The latter describes the geodesics motion of dust particles~\cite{Lim:2010yk}. 
In other words, we deal with PPF. Alternatively, one can calculate the pressure ${\cal P}$ and show that it equals to zero in the limit $H^2 \ll M^2$.
Consistently, the energy density 
\begin{equation}
\nonumber
\rho (t)=\frac{|\dot{\Psi}|^2}{2} +\frac{M^2|\Psi|^2}{2}  \; .
\end{equation}
redshifts with the scale factor as $\rho (t) \sim \frac{1}{a^3}$ in the same limit. In the opposite situation, when the Hubble rate is large compared to the mass $M$, the equation of state is that of the stiff matter~\cite{Li}, i.e., $\rho ={\cal P}$. We will 
not be interested in those early times, however.

We conclude that PPF is indeed 
reproduced from the massive complex scalar field upon tuning the initial conditions.

\subsection{Inhomogeneous evolution}


Let us now switch to the case of our primary interest---inhomogeneous evolution of PPF and the complex scalar. 
Our goal is to specify conditions, which bring together these two seemingly different models.  It is convenient to work with the complex field representation 
in terms of the normalized amplitude $\tilde{\lambda}$ and the phase $\tilde{\varphi}$ defined by Eq.~\eqref{complexdef}. The time derivative of the field $\Psi$ is then given by, 
\begin{equation}
\nonumber
\dot{\Psi}= \frac{\partial \ln \tilde{\lambda}}{\partial t} \Psi+i\cdot \frac{\partial  \tilde{\varphi}}{\partial t}  \Psi \; .
\end{equation}
We can express the time derivative of the phase $\tilde{\varphi}$ as follows, 
\begin{equation}
\nonumber 
 \frac{\partial  \tilde{\varphi}}{\partial t}=\mbox{Im} \frac{\dot{\Psi}}{\Psi} \; .
\end{equation}
The quantity $\frac{\partial \tilde{\varphi}}{\partial t}$ including its inhomogeneous perturbation is not arbitrary in the case of PPF but is defined by the constraint equation, i.e., 
\begin{equation}
\label{outofconstr}
 \frac{\partial \tilde{\varphi}}{\partial t}= \pm \sqrt{M^2+(\partial_i  \tilde{\varphi})^2} \; .
\end{equation}
In what follows we stick to the plus sign on the r.h.s. This corresponds to the PPF velocity defined as $v_i=-\partial_i  \varphi$. 
Therefore, the initial condition for the time derivative of the complex scalar cannot be arbitrary (if we are willing to reproduce PPF) but is fixed to be,
\begin{equation}
\label{PPFlike}
\mbox{Im} \frac{\dot{\Psi}}{\Psi}= \sqrt{M^2+(\partial_i \delta \tilde{\varphi})^2} \; .
\end{equation}
Let us argue that this condition is automatically satisfied in the limit of large $M$, i.e., $M \rightarrow \infty$.


Recall that the condition~\eqref{initial} should be obeyed in the homogeneous case. This fixes the generic solution for the complex scalar to be of the form, 
\begin{equation}
\label{interrel}
\Psi =\int dk \alpha (k) e^{ikx +i \sqrt{k^2+M^2}t}  \; .
\end{equation}
Using the latter, one writes for the time derivative of the field $\Psi$,
\begin{equation}
\label{dotpsiexact}
\dot{\Psi}= i \sqrt{-\partial^2_i +M^2} \; \Psi  \; ,
\end{equation}
The operator $\sqrt{-\partial^2_i +M^2} $ is defined by its Taylor expansion, 
\begin{equation}
\nonumber
\sqrt{-\partial^2_i +M^2} = M \left(1+\sum_n (-1)^n \alpha_n \frac{\partial^{2n}_i}{M^{2n}} \right) \; ,
\end{equation}
where $\alpha_n$ are the coefficients of the expansion. Their precise values will not be relevant for us. For simplicity, consider the case $n=1$. We have, 
\begin{equation}
\nonumber 
\partial^2_i  \Psi= \left(\frac{\partial^2_i  \tilde{\lambda}}{\tilde{\lambda}}+\frac{2i \partial_i \tilde{\lambda}}{\tilde{\lambda}}\partial_i  \tilde{\varphi} +i \partial^2_i  \tilde{\varphi} -(\partial_i  \tilde{\varphi})^2 \right) \Psi  \; . 
\end{equation}
The last term in brackets on the r.h.s. is the most relevant one, as it involves the second power of the phase $\tilde{\varphi}$. 
The latter is a large quantity. This follows from the chain of equalities $\partial_i  \tilde{\varphi} = M \partial_i  \varphi =- Mv$. 
Hence, for the fixed velocity $v$, the quantity $ \tilde{\varphi}$ grows as $M$, and the term $(\partial_i  \tilde{\varphi})^2$ indeed dominates 
in the large $M$ limit. At least, this is true, whenever the spatial distribution of the amplitude $\tilde{\lambda}$ and the phase $\tilde{\varphi}$ is sufficiently smooth, i.e., 
\begin{equation}
\label{conditions}
\left| \frac{\partial^2_i  \tilde{\lambda}}{\tilde{\lambda}} \right|  \sim L^{-2} \ll  M^2 v^2 \qquad \left| \frac{\partial_i  \tilde{\lambda}}{\tilde{\lambda}} \right| \sim L^{-1} \ll  Mv \qquad |\partial^2_i \tilde{\varphi}| \sim \frac{Mv}{L} \ll M^2 v^2 \; .
\end{equation}
Here $L$ is the characteristic scale of inhomogeneities in the amplitude $\tilde{\lambda}$ and the phase $\tilde{\varphi}$. If $M^{-1}$ is some microscopic scale, that condition is satisfied with an excess 
in most situations of interest in cosmology and astrophysics. Consequently, we get
\begin{equation}
\nonumber 
\partial^2_i \Psi =-(\partial_i \tilde{\varphi})^2 \Psi \qquad (M \rightarrow \infty) \; .
\end{equation}
The generalization to the case of arbitrary $n$ is straightforward, 
\begin{equation}
\label{interm}
(-1)^n \partial^{2n}_i \Psi= (\partial_i   \tilde{\varphi})^{2n} \Psi \qquad (M \rightarrow \infty)\; .
\end{equation}
We conclude that 
\begin{equation}
\nonumber
\sqrt{-\partial^2_i+M^2} \Psi =\sqrt{(\partial_i  \tilde{\varphi})^2+M^2} \Psi \qquad (M \rightarrow \infty)\; .
\end{equation}
Substituting this into Eq.~\eqref{dotpsiexact}, we see that Eq.~\eqref{PPFlike} is indeed satisfied in the large $M$ limit.  

To summarize:  The negative-frequency branch of the generic solution of the free complex scalar field reproduces PPF in the limit 
of large $M$, provided that the distribution of the fields $\tilde{\lambda}$ and $\tilde{\varphi}$ is sufficiently smooth in space, i.e., the inequalities~\eqref{conditions} are obeyed. 
That conclusion could be anticipated from the simpler considerations of the degree of freedom counting. Indeed, PPF is described by one degree of freedom and, hence, is solved by two initial conditions. 
Consistently, the negative-frequency branch of the field $\Psi$ is defined by one complex amplitude $\alpha (k)$, which is  once again fixed by two real functions on the initial Cauchy surface. 

As the inhomogeneities in the field $\tilde{\lambda}$ grow, the inequalities~\eqref{conditions} become progressively 
less accurate and so the correspondence between the complex scalar and PPF. The discrepancy gets particularly strong close to the times, 
when the caustic singularity is supposed to be formed. This is basically the mechanism of completing PPF by means of the complex scalar. 
Soon, we will give a support in favor of this picture. Before that, let us establish the connection with another closely related way of 
completing PPF.


\subsection{Non-relativistic limit: connection to Schroedinger equation}

 Since this point on and until the end of the Section, we switch to the non-relativistic limit. The 
solution for the complex scalar field then takes the form,  
\begin{equation}
\nonumber
\Psi =e^{iMt} \tilde{\Psi} \qquad \tilde{\Psi} =\int dk \alpha (k) e^{ikx +i \frac{k^2}{2M}t} \; ,
\end{equation}
where we readily dropped the positive-frequency part of the solution. It is straightforward to see that the 
function $\tilde{\Psi}$ satisfies the Schroedinger equation, 
\begin{equation}
\label{Schr}
i \frac{\partial \tilde{\Psi}}{\partial t}= \frac{\partial^2}{2M\partial x^2} \tilde{\Psi} \; .
\end{equation}
Of course, its appearance is not a surprise, as the Klein-Gordon equation has been originally designed as the relativistic completion of the 
Schroedinger equation. 

The possibility to complete PPF by means of the Schroedinger equation is quite well-known in the literature~\cite{Widrow:1993qq, Widrow:1996eq, Davies:1996kp, Coles:2002sj, Short:2006md, Short:2006me, Uhlemann:2014npa}. 
So, it is often used to model a collection of collisionless particles. 
This may be relevant for the study of the gravitational clustering,---a complicated process, which requires running cumbersome N-body simulations. 
There is also one realistic situation, when the Schroedinger equation arises as the genuine completion of PPF. 
This is the case of ultra-light axions~\cite{Hu:2000ke, Woo:2008nn, Hui:2016ltb}. Namely, when the de Broglie wavelength of the axion is larger 
than the distance between the particles, the description in terms of the wave function becomes more adequate.

To set a connection between PPF and the quantum mechanical wave function, one performs the so called 
Madelung transformation,  
\begin{equation}
\nonumber
\tilde{\Psi}=\frac{\lambda}{M} e^{i M\delta \varphi} \; ,
\end{equation}
where we made use of the 'non-canonical' amplitude $\lambda$ and the phase $\varphi$---most relevant for the case of PPF. In terms of $\lambda$ and $\varphi$, the Schroedinger equation can be equivalently written as the system of coupled equations, 
\begin{equation}
\label{Madelung1}
\frac{\partial {\bf v}}{\partial t}+({\bf v} \cdot {\bf \nabla}) {\bf v}=-\frac{1}{2M^2 } \cdot {\bf \nabla}\frac{\Delta \lambda}{ \lambda}
\end{equation}
and 
\begin{equation}
\label{Madelung2}
\frac{\partial  \lambda^2}{\partial t}+{\bf \nabla} ( \lambda^2 {\bf v})=0 \; .
\end{equation}
Obviously, the same equations could be obtained immediately from Eqs.~\eqref{lambdaeq} and~\eqref{varphieq} upon imposing the Newtonian limit. The term on the r.h.s. 
of Eq.~\eqref{Madelung1} is often called 'quantum pressure'. It relies on the spatial derivatives of the field $\lambda$. Hence, the quantum pressure 
is negligible, provided that the field $\lambda$ is distributed smoothly in space, i.e., when the following inequality is obeyed
\begin{equation}
\nonumber
\left | \frac{{\bf \nabla} \frac{\Delta  \lambda}{\lambda}}{M^2 ({\bf v} \cdot {\bf \nabla})  {\bf v}} \right | \sim \frac{1}{L^2 \cdot M^2 \cdot v^2} \ll 1\; .
\end{equation}
Not surprisingly, we have arrived at our condition~\eqref{conditions}. Once it is fulfilled, we result with the pressureless 
Euler equation describing the non-relativistic evolution of collisionless dust particles. 
This is known to be plagued by caustic singularities. At later times, when inhomogeneities of the field $\lambda$ grow, the quantum pressure cannot be ignored anymore, and one gets a chance to avoid instabilities.

While we omit metric perturbations in the present paper, including the gravitational potential $\Phi$ is straightforward. One makes the replacement,  
\begin{equation}
\nonumber
-\frac{\partial^2}{2M\partial x^2} \rightarrow -\frac{\partial^2}{2M\partial x^2}+M\Phi \; .
\end{equation}
At the level of Madelung equations, this amounts to adding the gradient $-{\bf \nabla} \Phi$ to the r.h.s. of 
Eq.~\eqref{Madelung1}.

\subsection{Simple example}
The canonical complex scalar is manifestly free of caustic singularities. 
Nevertheless, it is interesting to see, how the real caustics of PPF is reflected in the complete picture and elucidate the mechanism of the singularity 
avoidance. This we do in the present Subsection by considering a tractable example. 

 \begin{figure}[tb!]
\begin{center}
\includegraphics[width=\columnwidth]{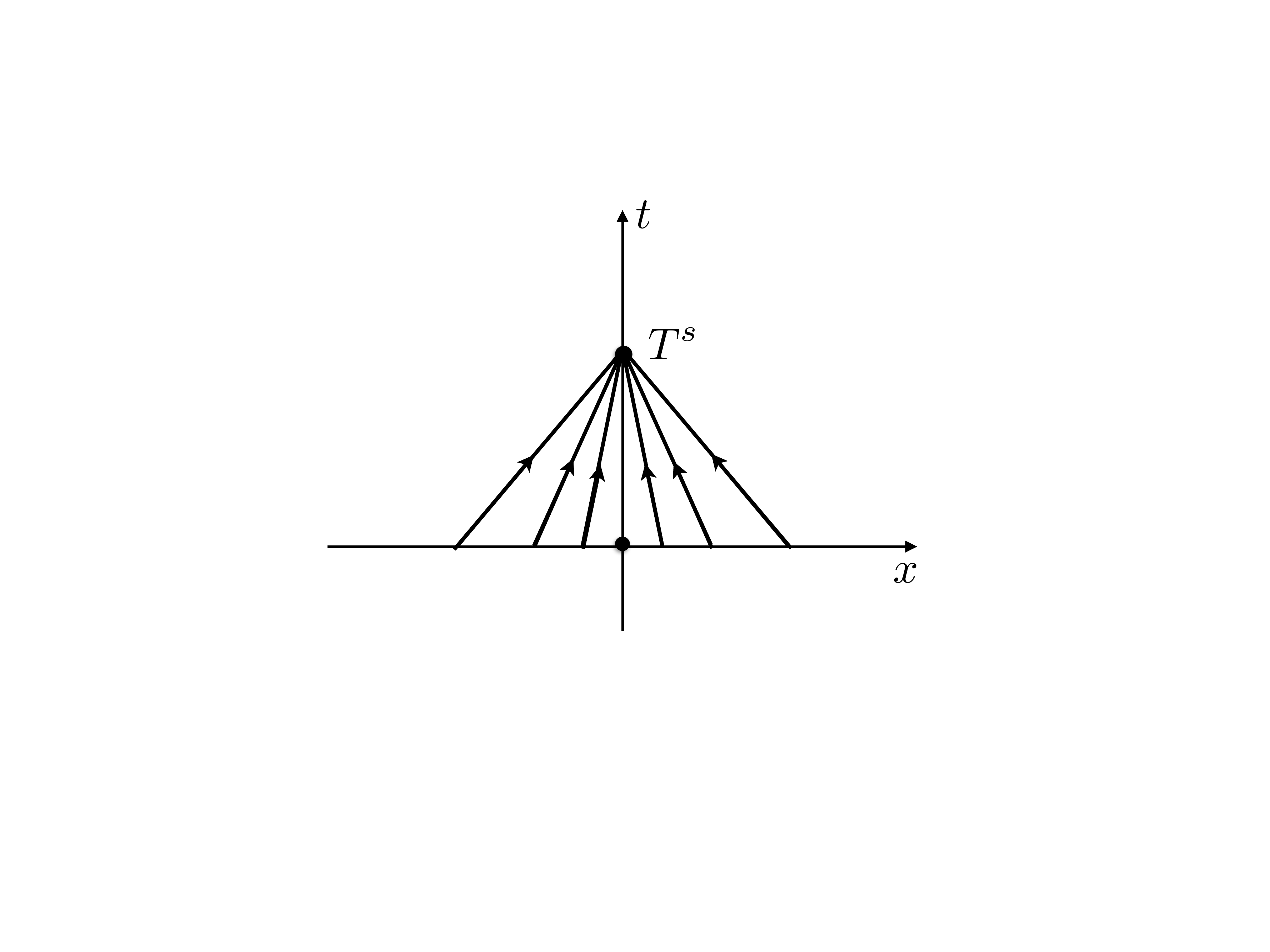}
\vspace{-5cm}
\caption{Characteristics of PPF for the  initial velocity profile~\eqref{initvelocity}. All characteristics cross at the same time $t=T^s$ forming the so called perfect caustics.}\label{perfect}
\end{center}
\end{figure}

Since the appearance of caustics is independent on the number of spatial dimensions, we can consider the 1-dimensional example. We also neglect the cosmological drag 
for simplicity, i.e., we set $a(t)=1$. We start with the following initial configuration of the complex scalar
\begin{equation}
\nonumber
\Psi =\tilde{\lambda} (x,t) e^{iMt +i \delta \tilde{\varphi} (x,t)}.
\end{equation}
Here we explicitly assumed the 'cosmological' background value for the normalized phase $\langle \tilde{\varphi} \rangle=Mt$. We choose 
sufficiently smooth initial conditions for the normalized amplitude $\tilde{\lambda}$ and the phase perturbation, 
\begin{equation}
\label{init}
\tilde{\lambda}= A \mbox{exp} \left(-\frac{x^2}{2L^2} \right) \qquad \delta \tilde{\varphi}\equiv \tilde{\varphi}-Mt=\frac{Bx^2}{2{L'}^2} \; .
\end{equation}
Here $A$ and $B$ are some arbitrary constants; the length scales $L$ and $L'$ characterize the size of initial inhomogeneities. 
The same choice of initial conditions was made in Ref.~\cite{Hui:2016ltb}, 
where the ultra-light axion in the Bose-Einstein condensate state was discussed. 
 That choice is convenient, because it corresponds to the integrable Gaussian profile for the field $\Psi$.

The velocity following from the initial distribution of the phase is given by, 
\begin{equation}
\label{initvelocity}
v=-\frac{d}{Mdx} \delta \tilde{\varphi} (x)  =-\frac{Bx}{M{L'}^2} \; .
\end{equation}
Such a velocity profile growing linearly with the coordinate results into the so called perfect caustics~\cite{ArkaniHamed:2005gu}. 
Employing for an instant the analogy with dust, all the particles fall into the center $(x=0)$ at the {\it same} time 
leading to the multivalued velocity nearby $x=0$. See Fig.~\ref{perfect}.

Before moving on, let us comment on the shortcoming of our profile~\eqref{initvelocity} choice. 
In the case of PPF, the Euler equation with the initial condition~\eqref{initvelocity} can be easily integrated out. The result reads, 
\begin{equation}
\label{veleuler}
v=-\frac{x}{T^{s}-t} \; ,
\end{equation}
where $T^s$ is the constant of integration defined from the initial condition~\eqref{initvelocity}, i.e., 
\begin{equation}
\nonumber
T^{s}=\frac{M{L'}^2}{B} \; .
\end{equation}
We see that at times $t \rightarrow T^s$, the velocity blows up at each point $x$. This is, however, not a physical 
singularity, as it stems from admitting infinite velocities in the initial distribution~\eqref{initvelocity}. 
For a realistic smooth distribution, the solution~\eqref{veleuler} gets modified 
$v \rightarrow -\frac{x}{T^s-t} +{\cal O} \left(\frac{x^3}{\left[T^s-t \right]^3}  \right)$. Hence, in the limit of interest, 
$t \rightarrow T^s$, it can be trusted only in the vicinity of the point $x=0$. To see the real singularity, 
one should instead consider the velocity divergence, i.e., 
\begin{equation}
\nonumber
\partial v (x=0)=-\frac{1}{T^s-t} \; .
\end{equation}
This is a trustworthy expression. It shows explicitly that the description in terms of PPF breaks down at the finite time $t=T^s$.

The solution for the complex scalar is given by, 
\begin{equation}
\label{inter}
\Psi =\int dk \alpha (k) e^{ikx +i\frac{k^2}{2M}t +iMt} \; .
\end{equation}
Writing it in this form, we explicitly assume the non-relativistic limit. It is straightforward to find the amplitude $\alpha (k)$ from the initial conditions for the field $\Psi$ (or $\tilde{\lambda}$ and $\tilde{\varphi}$). It reads 
\begin{equation}
\nonumber
\alpha (k)= \frac{A}{\sqrt{-2\pi i}} \sqrt{\frac{T}{M}} \cdot \mbox{exp} \left(-\frac{ik^2 T}{2M} \right) \; ,
\end{equation}
where we introduced the notation 
\begin{equation}
T\equiv T_1 -iT_2 = \frac{ML^2 {L'}^2}{B^2 L^4 +{L'}^4} \left(BL^2 -i {L'}^2 \right)\; .
\end{equation}
Note that in the limit $M \rightarrow \infty$, the time $T \rightarrow T^{s}$. On the other hand, 
for any large, but finite $M$, T is a complex quantity. This observation is at the core of solving 
the caustic singularity in the picture of the complex scalar. 

Substituting this amplitude into Eq.~\eqref{inter} and integrating over the momentum $k$, we get finally
\begin{equation}
\label{psifinal}
\Psi = \frac{A \cdot  e^{iMt}}{\sqrt{1-\frac{t}{T}}} \cdot \mbox{exp} \left(-\frac{iMx^2}{2(t-T)} \right)  \; .
\end{equation}
The velocity $v=-\partial_x \varphi$ is related to the field $\Psi$ by   
\begin{equation}
\nonumber
v=-\frac{\Psi_2 \partial_x \Psi_1-\Psi_1 \partial_x \Psi_2}{M\cdot |\Psi|^2} \; .
\end{equation}
From Eq.~\eqref{psifinal}, we get 
\begin{equation}
\label{velocitycomp}
v=-\frac{x(T_1-t)}{(T_1-t)^2+T^2_2} \; ,
\end{equation}
(cf. Appendix E of Ref.~\cite{Hui:2016ltb}). 
This expression as well as its derivatives is manifestly finite at all the times, as it should be. 
Notably, the velocity divergence, $\partial v$, is negative at $t<T_1$ and flips the sign at $t>T_1$. Using the analogy with particles, this 
corresponds to the situation, when particle trajectories tend to cross, but experience the repulsive force about the time $t \simeq T_1$. 
The repulsion is exactly due to the presence of the quantum pressure in Eq.~\eqref{Madelung1}, which is non-zero for any finite $M$.
On the other hand, in the limit $M \rightarrow \infty$, when $T_2 \rightarrow 0$, Eq.~\eqref{velocitycomp} reduces to the expression 
for the PPF velocity and the would be caustics appears. 

The same conclusion holds for the energy density $\lambda^2$ given by, 
\begin{equation}
\nonumber
\lambda^2 =\frac{A^2 M^2}{ \left| \sqrt{1-\frac{t}{T}}\right|^2} \cdot \mbox{exp} \left( -\frac{Mx^2 T_2}{[(t-T_1)^2+T^2_2]} \right) \; .
\end{equation}
The latter always remains finite contrary to the case of PPF. At the time $t=0$, thic correctly matches the initial condition~\eqref{init} for the field $\tilde{\lambda}$. 
This serves as a simple cross-check of our calculations.  

Despite multiple simplifications considered in the present Section, we assume that our example correctly reflects the real picture: 
the would-be collapse time is promoted to the complex quantity. Hence, the actual instability never occurs in the real time.

\section{k-essence}

According to the classification of Section~2, shift-symmetric k-essence scenarios correspond to the self-interacting complex scalar. The general analytic 
solution is not available in that case. Therefore, comparing k-essence and the complex field is a 
rather challenging task. This is still doable in the homogeneous case---the main focus of the present Section. Regarding the inhomogeneous 
evolution, we will be satisfied with translating the statements of Section~3 into the context of k-essence.


The equation of motion for the self-interacting complex field is given by
\begin{equation}
\label{eqk}
\ddot{\Psi} +3H\dot{\Psi}+2\frac{\partial V(|\Psi|)}{\partial |\Psi|^2} \Psi=0 \; .
\end{equation}
For an instant let us neglect the cosmic drag. The equation admits the simple oscillatory solution, 
\begin{equation}
\label{simpleosc}
\Psi =A e^{\pm i \omega t} \; ,
\end{equation}
where the frequency $\omega$ is given by 
\begin{equation}
\label{frequency}
\omega =\sqrt{2\frac{\partial V(|\Psi|)}{\partial |\Psi|^2}} \; .
\end{equation}
Namely, we pick the specific configuration of the field $\Psi$ described by the particular 
frequency dependence. See Fig.~\ref{potentials}.

Switching to the realistic case of the expanding Universe is straightforward. Though $|\Psi|$ is not 
constant anymore, it is a slowly varying function of time. Therefore, we can solve Eq.~\eqref{eqk} in the WKB approximation, 
\begin{equation}
\nonumber
\Psi=\frac{A}{a^{3/2} \sqrt{2\omega}} \cdot e^{\pm i \int \omega (t) dt} \; .
\end{equation}
To find the time dependence of the frequency $\omega$, let us take the absolute value squared of the left and right hand sides 
of the solution above,
\begin{equation}
2 \sqrt{2 \frac{\partial V}{\partial |\Psi|^2}} |\Psi|^2= \frac{A^2}{a^3} \; .
\end{equation}
Using this and given the potential $V$, one can find $|\Psi|$. Plugging the result into Eq.~\eqref{frequency}, we obtain the frequency $\omega$ time dependence. 

As a concrete example, consider the potential $V(\Psi) = M^4 |\Psi|^4/4\Lambda^4$. In that case, the frequency is given by
\begin{equation}
\label{ansatzrad}
\omega =\frac{1}{a} \left(\frac{A^2 M^4}{2\Lambda^4} \right)^{1/3}  \; ,
\end{equation}
and the solution for the field $\Psi$ can be written as follows, 
\begin{equation}
\nonumber
\Psi \propto \frac{1}{a} \cdot e^{ \pm i\int \omega dt } \; .
\end{equation}
Such a profile of the field $\Psi$ corresponds to radiation. To show this, consider the energy density
\begin{equation}
\nonumber
\rho (t)=\frac{1}{2}| \dot{\Psi} |^2+V(\Psi) \; .
\end{equation}
For the quartic potential $V$, it redshifts as
\begin{equation}
\nonumber
\rho (t) \propto \frac{1}{a^4},
\end{equation}
what proves the statement.
This conclusion perfectly matches the result obtained in the k-essence scenario. Indeed, the quartic potential stands for the Lagrangian of the form ${\cal L} (X)= X^2$. 
This Lagrangian effectively describes the radiation, as it should be. 

The inhomogeneous evolution of the canonical complex scalar is obviously caustic free. Still, it is unclear, 
if it reproduces k-essence models in the limit of large $M$. The issue is complicated due to the presence of the self-interacting potential. 
Therefore, we will formulate our conclusions by exploiting the analogy with PPF. With inhomogeneities included, the free scalar retains the PPF-like behaviour. 
We assume that the same works for k-essence. Hence, our conjecture: {\it The self-interacting complex scalar reproduces k-essence, given that it has a fixed frequency 
dependence, namely its homogeneous profile satisfies}
\begin{equation}
\nonumber
\Psi=\frac{A}{a^{3/2} \sqrt{2\omega}} \cdot e^{\pm  i \int \omega (t) dt}  \qquad t \rightarrow 0 \; .
\end{equation}
The correspondence, we assume, holds until the times, when singularities are supposed to be formed. 
As inhomogeneities of the fields $\tilde{\lambda}$ and $\tilde{\varphi}$ grow, the discrepancy between 
k-essence and the complex scalar becomes large. In particular, while the description 
in terms of k-essence breaks down at some point, no actual singularity occurs in the complete picture.

{\it Sub-/superluminality} Let us now comment on another pesky property of the k-essence: its perturbations 
exhibit the sub-/superluminality. The sound speed squared of k-essence perturbations propagating in the 
preferred background is given by~\cite{Garriga:1999vw} 
\begin{equation}
\nonumber 
c^2_s =\left(1+2X \frac{{\cal L}_{XX}}{{\cal L}_{X}} \right)^{-1} \; .
\end{equation}
Taking again the Lagrangian ${\cal L} \propto X^2$, we reproduce the standard result $c^2_s =\frac{1}{3}$, as it should be in the case of radiation. 

For positive $X$ and ${\cal L}_{X}$, the sign of ${\cal L}_{XX}$ defines, if the 
sound speed is subluminal (${\cal L}_{XX}>0$) or superluminal (${\cal L}_{XX}<0$). This property can be easily explained in the complete picture. Components of the complex scalar are 
manifestly luminal in all backgrounds. On the other hand, switching to the 'inconvenient' variables of the amplitude $\tilde{\lambda}$ and the 
phase $\tilde{\varphi}$, makes the property of luminality less transparent. The phase $\tilde{\varphi}$ has a non-canonical kinetic term, which relies on the background value of the amplitude $\tilde{\lambda}$. 
Hence, for generic backgrounds, the emergence of sub-/superluminality is inevitable. We get back to the conventional luminality upon switching to the canonical variables. 

\section{Conclusions}

In the present work, we pursued the unified completion of pressureless perfect fluid (PPF) and the shift-symmetric k-essence scenarios. 
In Section~2, we showed that they belong to the same class of models involving two scalars. This class can be easily completed by means 
of the unique complex field. We derived our main conclusions in Section~3 by 
exploiting the tractable example of the free massive scalar, which was our main reference 
point. Despite the simplicity, this describes the physically interesting model---PPF. In Section~4, we generalized the discussion to the case of k-essence.

One of our conclusions concerns the mechanism of caustic singularity avoidance. We 
observed that the PPF collapse time is promoted to the complex number in the complete picture. 
Hence, the real time evolution always remains smooth in the case of the canonical scalar field, as it should be. 
This observation may have applications beyond the scope of the 
present research. Indeed, in the non-relativistic limit, the fixed frequency branch of the complex scalar reduces 
to the quantum mechanical wave function obeying the Schroedinger equation. The latter is often used to model collisionless particles without 
using N-body simulations. Finally, the ultra-light axion is described by the scalar field at sufficiently small 
scales (still comparable with the size of halos). In all those cases, the mechanism of caustic avoidance discussed in Subsection 3.4 is applied. 

Finally, we would like to point out several open issues. 
First, in the inflationary Universe the amplitude of the 
scalar $\Psi$ rapidly tends to zero with a high accuracy. 
In this situation, the complex scalar represents just a collection 
of heavy particles above the trivial vacuum. Instead, we are interested in the non-trivial classical configuration of the complex field shown in Fig.~\ref{potentials}. However, this non-trivial configuration 
can be generated, if we admit a slight 
breaking of the shift-symmetry (or, equivalently, $U (1)$-symmetry). The latter can be achieved, 
e.g., by coupling the phase $\tilde{\varphi}$ to the matter fields. 
More worrisome is our assumption about tuned initial conditions 
for the complex scalar. Recall, that the latter should have a particular frequency 
dependence. This may strongly constrain the mechanism 
of generating the field $\Psi$ in the early Universe. We plan to get 
back to these issues in the future.

{\bf Acknowledgments}
We are indebted to Gilles Esposito-Farese for useful discussions. EB acknowledges support from the research program, Programme national de cosmologie et Galaxies of the CNRS/INSU, France, from the project DEFI InFIniTI 2017 and from Russian Foundation for Basic Research Grant No. RFBR 15-02-05038.

\end{document}